\documentclass[12pt,a4paper]{article}
\newcommand{\beq}{\begin{equation}}
\newcommand{\eeq}{\end{equation}}
\newcommand{\beqy}{\begin{eqnarray}}
\newcommand{\eeqy}{\end{eqnarray}}

\newcommand{\as}{\alpha_s}

\newcommand{\ee}{e^+e^-}

\begin{document}
\title{On the cross-section peaks for a heavy quark bound state}
\author{ Nicola~Fabiano \footnote{nicola.fabiano@pg.infn.it} \\
Istituto Nazionale di Fisica Nucleare, \\
Sezione di Perugia, Via A. Pascoli I-06123, Perugia, Italy}

\date{}
\maketitle
\abstract{We discuss the values of resonance peaks of  the cross--section of a 
heavy quark  bound state obtained by means of a Green function method 
applied to a Coulombic model and compare the result to the $\Upsilon$ 
and $J/\psi$ data.}
\section{Introduction}
\label{sec:intro}

The total cross--section of a heavy quark bound state
resonance in an $e^+e^-$ annihilation
is described by the Breit--Wigner formula
\beq
\sigma(\sqrt{s}) = \frac{3 \pi}{M^2} \frac{\Gamma_{ee}\Gamma}{(\sqrt{s}-M)^2 + \Gamma^2/4}
\label{eq:bw}
\eeq
where $M$ is the mass of the resonance, $\sqrt{s}$ the centre of mass 
energy, $\Gamma$
the total width and $\Gamma_{ee}$ the decay width into electrons.

For a bound state however a better analytical description of its total 
cross--section is given by the imaginary part of the Green function of the bound
state itself~\cite{Fadin:1987wz,Fadin:1988fn,Meixner:1933}. 
The basic idea is to consider the Schr\"odinger equation of the bound state
and compute its Green function
\beq
(\mathbf{H}-E) \mathcal{G}(\mathbf{x},\mathbf{y},E) = \delta(\mathbf{x}-\mathbf{y})
\label{eq:es}
\eeq
where $\mathbf{H}$ is the Hamiltonian of the system
\beq
\mathbf{H} = - \frac{\nabla_{\mathbf{x}}^2}{2m} +V(\mathbf{x}) .
\label{eq:H}
\eeq
The imaginary part of the derivative of the Green function given 
by~(\ref{eq:es}) taken at the origin is proportional to the 
cross--section at threshold. 
The finite width of the state is taken into account by the substitution 
$E \to E +i \Gamma$, $E$ being
the energy offset from $2m$ threshold. 
So for the total cross--section we obtain
\beq
\sigma(E,\Gamma) \sim \sigma_{\ee} \Im \mathcal{G}(0,0,E+i\Gamma)
\label{eq:sigma}
\eeq
where $\sigma_{\ee}$ accounts for the $\ee$ part of the process and
depends on the energy solely by the function
$1/s$ if the energy of the process is below the $Z$ threshold. That is, this
formula involves only the photon propagator and not the $Z$ exchange together
with its interference effect.
The other factors
depend on the particular process taken into account and are not 
universal~\cite{Fabiano:2001cw,Penin:1998mx}. We have therefore 
factored eq.~(\ref{eq:sigma}) in a product of terms, the first one 
for the $\ee$ process and the second one for the hadronic part of cross--section.

\section{The Green function method}
\label{sec:method}
The next step is to compute the Green function for a realistic model. The
Coulombic potential
\beq
V(r) = - \frac{4}{3} \frac{\as}{r}
\label{eq:coulomb}
\eeq
where $r=|\mathbf{x}|$ provides an integrable system which is quite realistic
for a heavy quark bound state provided the QCD energy  scale of $\as$ is set to
the inverse of Born radius, $r_B=2/(3 m\as^2)$, 
see ~\cite{Fabiano:2001cw} and references therein. In this case we are able
to provide a fully analytical solution for the bound state energy levels,
namely
\beq
E_n = - \frac{4 }{9} \frac{m\as^2}{n^2}
\label{eq:ebound}
\eeq

The solution to problem~(\ref{eq:es}) in the case of $S$ wave is given 
in~\cite{Penin:1998mx} with a slight notation change

\beqy 
\mathcal{G}(0,0,E+i\Gamma) &= &\frac{m}{4 \pi} \left [-2 \lambda \left (
 \left (\frac{k}{2 \lambda} \right) + \right . \right . \nonumber \\ 
 +\log \left( \frac{k}{\mu} \right ) &  +&\psi(1-\nu)+
2 \gamma-1 \left . \left . \right ) \right ]
\label{eq:green1s}
\eeqy
where
$ k = -\sqrt{m (E+i \Gamma)}$, $ \lambda = 2 \as m/3$ and the
wavenumber $ \nu =\lambda/k $, $E=\sqrt{s}-2m$.
The $\psi$ is the logarithmic derivative of Euler's Gamma function $\Gamma(x)$,
$\gamma \simeq 0.57721$ is Euler's constant
and $\mu$ is a soft scale that cancels out in the determination of physical observables.

By inspection of eq.~(\ref{eq:green1s}) we see that the main contribution for
energies $E$ close to the one of
bound states~(\ref{eq:ebound}) is given by the $\psi$
function, which has simple poles for negative integers $-n$ in the 
complex plane:
\beqy
\psi(z) &=& -\frac{1}{z+n} + \psi(n+1) + \nonumber \\
\sum_{k=1}^{\infty} \left [ \frac{\psi^{(k)}(1)}{k!} \right . &+&
\left . \zeta(k+1)-\zeta(k+1,n+1) \right ] (z+n)^k
\label{eq:psising}
\eeqy
for $z \to -n$, where $\psi^{(k)}(z)$ is the $k$-th derivative of the 
$\psi(z)$ function with respect to $z$, the so--called polygamma function; 
$\zeta(z) = \sum_{k=1}^{\infty} k^{-z} $ 
is the Riemann zeta function, while 
$\zeta(z,q) = \sum_{k=1}^{\infty} (k+q)^{-z} $  is the Hurwitz zeta 
function~\cite{sondow:2006}.

\section{Results of the Coulombic model for $\Upsilon$ state}
\label{sec:results}
As previously discussed in sec.~\ref{sec:intro} the cross--section of a heavy
quark resonance below $Z$ mass could be written as
\beq
\sigma(E,\Gamma) \sim \frac{1}{s} \Im \mathcal{G}(0,0,E+i\Gamma)
\label{eq:sigmauniv}
\eeq
where in our particular model of Coulombic interaction $\mathcal{G}$
is given explicitly by formula~(\ref{eq:green1s}).

With~(\ref{eq:sigmauniv}) we are going to compute the ratio of the $2S$
and $1S$ resonance peaks respectively, which should be independent upon
the particular bound state chosen for this evaluation. In fact one obtains
the following expression
\beq
\frac{\sigma(E_{2S},\Gamma_{2S})}{\sigma(E_{1S},\Gamma_{1S})}=
\frac{M_{1S}^2}{M_{2S}^2} \times
\frac{\Im \mathcal{G}(0,0,E_{2S}+i\Gamma_{2S})}{\Im
\mathcal{G}(0,0,E_{1S}+i\Gamma_{1S})}
\label{eq:sigmaratio}
\eeq
which will be renamed as $\sigma(2)/\sigma(1)$ for sake of brevity. The first
term of RHS of eq.~(\ref{eq:sigmaratio}) is close to 1, and from 
eq.~(\ref{eq:ebound}) the mass of the bound state is given by
\beq
M_n = 2 m +E_n = 2m \left ( 1-\frac{2}{9} \frac{\as^2}{n^2} \right )
\label{eq:massbound}
\eeq
so that the ratio $M^2_{2S}/M^2_{1S}$ is given by 
$1+\as^2/3+ \mathcal{O}(\as^4)$.

Defining a suitable variable depending on a generic width value $t= \Gamma/E_1$,
with $E_1$ given by eq.~(\ref{eq:ebound}) for $n=1$, we obtain the fairly 
elegant expression
\beqy
\frac{\sigma(2)}{\sigma(1)} &=& \frac{1}{8}+
\frac{\left( 42\,{\pi }^{2}+425\right) \,{t}^{2}}{128}- \nonumber \\ 
&-&\frac{\left
( 147384\,\zeta\left( 3\right) +1134\,{\pi}^{4}+11096\,{\pi }^{2}+36545\right) 
\,{t}^{4}}{3072}+ \nonumber \\
 +\mathcal{O}(t^6)
\label{eq:sigmataylor}
\eeqy
which would suggest a value close to $1/8$ for this ratio. Unfortunately 
the~(\ref{eq:sigmataylor}) is a very slowly convergent series as its
numeric evaluation shows:
\beq
\frac{\sigma(2)}{\sigma(1)} = 0.125 + 6.559\,{t}^{2}-48.877\,{t}^{4}
+\mathcal{O}(t^6)
\label{eq:sigmataylornum}
\eeq
that is, the coefficient of $\mathcal{O}(t^k)$ term grows more than $k!$~,
suggesting to evaluate the result with the full 
expression of~(\ref{eq:green1s}) for the Green function.

We will make use now of this method with $\Upsilon$ resonances data.
This $b\overline{b}$ bound state is the ideal candidate for this method: it is
rather heavy, $m_b \sim 5$~GeV, so that it should allow us to neglect the
confining linear term proportional to $r$ in the QCD 
potential~\cite{Fabiano:1993vx,Fabiano:1994cz,Fabiano:1997xh}. 

For the $\Upsilon(1S)$ PDG data~\cite{Yao:2006px} give us:
\beqy
M_{\Upsilon}(1S) &=& 9.46030 \pm 0.00026 \textrm{~GeV~,~} \nonumber \\
 \Gamma(1S)& =& 54.02 \pm 1.25
\textrm{~keV~,}  \nonumber \\
\Gamma_{ee}(1S) &= &1.340 \pm 0.018\textrm{~keV~;}
\label{eq:ups1s}
\eeqy
while for the $\Upsilon(2S)$ one reads
\beqy
M_{\Upsilon}(2S) &=& 10.02326 \pm 0.00031  \textrm{~GeV~,~} \nonumber \\
\Gamma(2S) &= &31.98 \pm 2.63
\textrm{~keV~,}  \nonumber \\
\Gamma_{ee}(2S) &= &0.612 \pm 0.011\textrm{~keV~.}
\label{eq:ups2s}
\eeqy

Plugging all those data into the Coulombic model~(\ref{eq:coulomb}) and
eq.~(\ref{eq:sigmaratio}) we compute the ratio of the first two peaks for the 
$\Upsilon$ resonance obtaining the value

\beq
\frac{\sigma(2)}{\sigma(1)} = 0.211 \pm 0.024~~.
\label{eq:ratiocalc}
\eeq
This result is very sensitive to the exact total width value, and depends much
less on both the mass of the state and the exact $\as$ coupling of the Coulombic
model.
It differs from first term of the series expansion~(\ref{eq:sigmataylor})
by approximately a factor of 2 proving its slow convergence despite the 
smallness of its parameter $t$, of the order of $10^{-4}$.

\section{Comparison with data}
\label{sec:comparison}
We are ready now to compare our results with the $\Upsilon$ measurements. From
eq.~(\ref{eq:bw}) the maximal value of the Breit--Wigner
resonance cross--section is given by the expression
\beq
\sigma_{max} = \frac{12 \pi}{M^2} \frac{\Gamma_{ee}}{\Gamma}~.
\label{eq:bwmax}
\eeq
Using again the experimental data of eqs.~(\ref{eq:ups1s}),~(\ref{eq:ups2s})
we have the following value for the measured ratio:
\beq
\left ( \frac{\sigma(2)}{\sigma(1)} \right )_{exp} = 0.685 \pm 0.094~~,
\label{eq:ratiobw}
\eeq
which is quite sensitive to the exact determination of the widths.
It is clear that this value doesn't agree with~(\ref{eq:ratiocalc}); some
\emph{caveats} are in order here.

According to~\cite{Yao:2006px,Rosner:2005eu}, because of ISR (initial state
radiation) and Beamstrahlung effect the observed line shape is not simply
given by eq.~(\ref{eq:bw}) but a convolution of the $\Upsilon$ width
of $\mathcal{O}$(keV), of ISR and of beam energy spread 
of $\mathcal{O}$(MeV). The $\Gamma_{ee}$ value cannot therefore be directly
measured, but is calculated from the production cross--section of 
$\Upsilon$ integrated over the incoming $e^+e^-$ energies, that is
\beq
\int \sigma(e^+e^- \to \Upsilon) dE
\label{eq:sigmint}
\eeq
The integral itself however is again calculated from the Breit--Wigner
formula~(\ref{eq:bw}) in a bootstrap fashion, thus leading to a heavily model
dependent result of the cross--section shape.

The Coulombic model works well even at lower than $\Upsilon$
scales~\cite{Fabiano:2002bw}, there could be however some relativistic 
non negligible effect in this case (recall from sec.~\ref{sec:intro}
that this whole method is non--relativistic). From the virial theorem
applied to a Coulombic model we obtain the relation between the kinetic and
potential energies average
$ -\langle V \rangle/2 = \langle T \rangle$ that leads to a speed estimate of
the component quark inside the meson
\beq
\langle v^2 \rangle = \frac{8}{9} \, \as^2
\label{eq:speed}
\eeq
This brings a $\gamma$ relativistic correction of about 8\% at 
$\Upsilon$ scale that could change the computed ratio.

\section{Further estimates: the $J/\psi$ case}
\label{sec:comparisonjpsi}
To shed some more light on this discrepancy we have described
it could be useful to compare our
results to the one obtained from $J/\psi$ data. This $\overline{c}c$ 
bound state is less ideal than the former $\Upsilon$ for our purpose, 
as it is lighter ($m_c \sim 1.5$ GeV), thus linear confining terms for the
potential as well as relativistic corrections could have larger effects.
From PDG data for $J/\psi$~\cite{Yao:2006px} we read:
\beqy
M_{J/\psi}(1S) &=& 3096.916 \pm 0.011  \textrm{~MeV~,~} \nonumber \\
 \Gamma(1S)& =& 93.2 \pm 2.1
\textrm{~keV~,}  \nonumber \\
\Gamma_{ee}(1S) &= &5.55 \pm 0.14\textrm{~keV~;}
\label{eq:jpsi1s}
\eeqy
while for the $\psi(2S)$ one reads
\beqy
M_{\psi(2S)}(2S) &=& 3686.09 \pm 0.04  \textrm{~MeV~,~} \nonumber \\
\Gamma(2S) &= &317 \pm 9
\textrm{~keV~,}  \nonumber \\
\Gamma_{ee}(2S) &= &2.38 \pm 0.04\textrm{~keV~.}
\label{eq:jpsi2s}
\eeqy
Proceeding like described in sec.~\ref{sec:comparison} we obtain for the 
$J/\psi$ peak measured values
\beq
\left ( \frac{\sigma(2)}{\sigma(1)} \right )_{exp} = 0.0890 \pm 0.0083~~.
\label{eq:ratiobwjpsi}
\eeq
The Green function approach method detailed in sections~\ref{sec:method} 
and~\ref{sec:results} gives us 
\beq
\frac{\sigma(2)}{\sigma(1)}=0.0368 \pm 0.0019  ~~.
\label{eq:ratiocalcjpsi}
\eeq
A comparison of eq.~(\ref{eq:ratiocalcjpsi}) with eq.~(\ref{eq:ratiobwjpsi})
shows that the result for the $J/\psi$ case is slightly worse than the one
for $\Upsilon$ as expected. In fact for the former case the two central 
values differ for more than $27\sigma_G$, while in the latter the difference is 
less pronounced, above $19\sigma_G$
(here $\sigma_G$ refers to the error given by the Green function procedure).
An estimate of the relativistic $\gamma$ correction for $J/\psi$ done in the 
same fashion of eq.~(\ref{eq:speed}) gives us a result of about $38\%$.

This $\gamma$ correction together with the absence of a linear term in the
potential could account for the larger difference seen at a lower energy scale.

We must stress again that this method of the Green function for a
Coulombic potential of a  Schr\"odinger equation 
furnishes us with an \emph{exact} analytical solution given
by eq.~(\ref{eq:green1s}) only in the case of this particular interaction,
and only for a non--relativistic system. 
The addition of further correction terms to the
potential of eq.~(\ref{eq:coulomb}) like a linear confining term $r$ or a
relativistic correction of order $v^2$ would spoil the integrability of this
problem and thus the possibility of full control over solutions.
For instance, a Hamiltonian with a funnel potential like 
$V(r) = -4\as/(3r) +ar$ or a Hamiltonian
with relativistic correction $H=p^2/(2m)-p^4/(8m^3)+V(r)$ are not
exactly solvable, and even a perturbative approach is unfit in our
case as corrections to the original $H$ of eqs.~(\ref{eq:H}) 
and~(\ref{eq:coulomb})  are not small.
Those systems would call for a purely numerical search of results which 
is beyond the scope of this paper.

\section{Conclusions}
\label{sec:end}
We have compared the ratio of the first two cross--section resonance 
$\Upsilon$ peaks computed from a QCD model and the one obtained from
the experiments, given respectively by $0.211 \pm 0.024$ and
$0.685 \pm 0.094$. 
Another comparison is done for the $J/\psi$ case, giving a calculated
value of $0.0890 \pm 0.0083$
and an experimental one of $0.0368 \pm 0.0019$ for peaks ratio.
Albeit the two results do not agree with each other, it is necessary
to consider that the theoretical model could need some relativistic corrections.
On the other hand, the experiments do not give a direct measurement of the
peaks, but rather depend  on the model used to  
evaluate the cross--section. 

Therefore it should be possible to compare again the two results using
more refined methods from the theoretical model side and from  the 
measure technique as well.\\

\textbf{Acknowledgments} \\
The work of N.~F. was supported by the {\scshape Fondazione Cassa di Risparmio 
di Spoleto}. The author wishes to thank Y.~Srivastava for constructive
criticism.

\end{document}